\documentclass[10pt]{article}
\usepackage{latexsym}
\usepackage{amssymb}
\usepackage{amsmath}
\usepackage{amscd}
\usepackage{amsthm}
\usepackage[left=2cm,top=2.5cm,right=2.5cm,bottom=1.5cm]{geometry}
\usepackage[dvips]{graphicx}
\usepackage{epstopdf}
\usepackage{hyperref}
\begin{document}
\begin{center}
\large{\bf{Interacting Viscous Dark Energy in Bianchi Type-I Universe}} \\
\vspace{10mm}
\normalsize{Hassan Amirhashchi} \\
\vspace{5mm}
\normalsize{Department of Physics, Mahshahr Branch, Islamic Azad University, Mahshahr, Iran \\
\vspace{2mm}
E-mail: hashchi@yahoo.com} \\
\end{center}
\vspace{10mm}
%\date{}
%\maketitle
\begin{abstract}
A solution to the coincidence and Big Rip problems on the bases of an anisotropic space-time is proposed. To do so, we study the interaction between viscous dark energy and dark matter in the scope of the Bianchi type-I Universe. We parameterize the viscosity and the interaction between the two fluids by constants $\zeta_{0}$ and $\sigma$ respectively. A detailed investigation on the cosmological implications of this parametrization has been made. We have also performed a geometrical diagnostic by using the statefinder pairs $\{s, r\}$ and $\{q, r\}$ in order to differentiate between different dark energy models. Moreover, we fit the coupling parameter $\sigma$ as well as the Hubble's parameter $H_{0}$ of our model by minimizing the $\chi^{2}$ through the age differential method, involving a direct measurement of $H$.
\end{abstract}

%\newpage
%%%%%%%%%%%%%%%%%%%%%%%%%%%%%%%%%%%%%%%%%%%%%%%%%%%%%%%%%%%%%%%%%%%%%%%%%%%%%%%%%%%%%%%%%%%%%%%%%%%%%
%%%%%%%%%%%%%%%%%%%%%%%%%%%%%%%%%%%% SECTION 1  %%%%%%%%%%%%%%%%%%%%%%%%%%%%%%%%%%%%%%%%%%%%%%%%%%%%%
\section{Introduction}
The fact that we live in an accelerating expanding Universe is a well established reality today. The direct evidence of such an accelerating expansion of the Universe first published by the High-z Supernova Search Teams (Riess et al. 1998; Perlmutter et al. 1999) in 1998 and early 1999. Later on, some other observations such as measurements of cosmic microwave background (Miller et al. 1999; Bernardis et al. 2000; Halverson et al. 2002; Benoit et al. 2003; Mason et al. 2003) and the galaxy power spectrum (Spergal et al. 2003; Tegmark et al. 2004; Page et al. 2003) indicate that the expansion of our Universe is speeding up currently. These observations strongly suggest that we live in a nearly spatially flat
Universe with approximately $1/3$ of the critical energy density being in non-relativistic matter and $\sim 2/3$ in a smooth component with high negative pressure called dark energy (DE). The concept of dark energy refers to a kind of exotic energy with negative pressure which has been proposed for understanding the current accelerating expansion of the universe. Although, dark energy has very significant effect on the evolution of the universe at large scales and the ultimate fate of our universe will be determined by this unknown component, our knowledge about its nature and properties is still very limited. We only know the following well established information about dark energy (i) it is an exotic energy with negative pressure which is caused the present cosmic acceleration; (ii) it is non-clustering and spatially homogeneous; and (iii) it was small at the early times while dominates the present universe (at $z\sim 0.5$).\\

The study of dark energy is possible through its equation of state (EoS) parameter $\omega^{X} = p^{X}/\rho^{X}$. Here, $p^{X}$ and $\rho^{X}$ are the pressure and the energy density of DE respectively. Unfortunately, the current value of the dark energy EoS parameter is not known to us yet.
This is why so far many candidates have been proposed for dark energy. The first candidate for dark energy which can simply explain the observed cosmic acceleration is a cosmological constant (Weinberg 1989; Carroll 2001; Padmanabhan 2003; Peebles and Ratra 2003) with $\omega^{X}=-1$. However some fundamental theoretical problems, such as the fine-tuning problem and the coincidence problem associated with cosmological constant lead cosmologists to think about another candidates such as quintessence ($-1<\omega^{X}<-\frac{1}{3}$) (Wetterich 1988; Ratra and Peebles 1988), phantom ($\omega^{X}<-1$) (Caldwell 2002), quintom ($\omega^{X}<-\frac{1}{3}$) (Feng et al. 2005), Chaplygin gas as well as generalized Chaplygin gas models (Srivastava 2005; Bertolami et al. 2004; Bento et al. 2002; Alam et al. 2003), and etc. Quintessence dark energy models introduce a scalar field $\phi$ that is minimally coupled to gravity. Although, a possible solution to the cosmic coincidence problem my be provided by the attractor version of quintessence, but as dark energy, quintessence field with $\omega>-1$ is not in agreement with the recent observations (Hinshaw et al. 2009; Komatsu et al. 2009; Copeland et al. 2006; Perivolaropoulos 2006) which indicate that $\omega^{X}<-1$ is allowed at $68\%$ confidence level. Since the present observational data do not exclude the possibility of $\omega^{X}$ to be less than $-1$, a new class of scalar field models with negative kinetic energy, known as the phantom field models (Caldwell 2002) have been introduced. This scenario also would encounter two essential problems namely the Big Rip (Caldwell et al. 2003; Nesseris and Perivolaropoulos 2004) of the universe and the ultraviolet quantum instabilities problem (Carroll et al. 2003). Since the models with a transition from $\omega^{X}>-1$ to $\omega^{X}<-1$ are mildly favored by recent observations, a new scenario of dark energy dubbed Quintom was proposed (Feng et al. 2005). This model can be considered as a combination of quintessence and phantom dark energy models which can fulfill the transition from $\omega^{X}>-1$ to $\omega^{X}<-1$ and {\it vice versa}.\\

Recent astrophysical observational evidences such as the large entropy per baryon and the remarkable degree of the isotropy of cosmic microwave
background radiation (CMBR) indicate that the cosmic media is not a perfect fluid (Jaffe et al. 2005). These observed physical phenomena reveal the importance of analysis of the dissipative effects in cosmology. It is believed that the occurrence of the big rip can be avoided in dissipative dark energy models in which the negative pressure is an effective bulk viscosity (McInnes 2002; Barrow 2004). Zimdahl et al. (2001) and Balakin et al. (2003) have studied the role of viscous pressure as an agent that drives the present acceleration of the universe. Padmanabhan and Chitre (1987) have also investigated the possibility of a viscosity dominated late epoch of the Universe with accelerated expansion.
The effect of bulk viscosity on the background expansion of the universe has been studied from different points of view (Cataldo et al. 2005; Brevik ans Gorbunova 2005; Szydlowski and Hrycyna 2007; Singh 2008; Feng and Zhou 2009; Oliver et al. 2011; Amirhashchi 2013).
It is worth to mention that the general theory of dissipation in relativistic imperfect fluid was first suggested by Eckart (1940), Landau and Lifshitz (1987). \\

It is believed that an energy transition from dark energy to dark matter could lead to a solution to the coincidence problem in cosmology (Cimento et al. 2003; Dalal et al. 2001; Jamil and Rashid 2008; Jamil and Rashid 2009). In other word, by considering a coupling between dark energy and dark matter, we can explain why the energy densities of DE and DM are nearly equal today. It is worth to mention that the recent observations (Bertolami et al. 2007; Le Delliou et al. 2007; Berger and Shojaei 2006) provide the evidence for the possibility of such an interaction between these two dark components. From theoretical point of view, the possibility of the interaction between DE and DM have been widely investigated in the literatures. Zhang (2005a, 2005b), Zimdahl and Pavon (2004), Setare (2007a, 2007b, 2007c, 2009a, 2009b), Setare et al. (2009, 2011), Banijamali et al. (2009), Sheykhi and Setare (2010, 2011), Pradhan et al (2011a, 2011b), Saha et al. (2012), and Amirhashchi et al. (2011a, 2011b, 2011c) have investigated dark energy in different contexts.
Viscous dark energy and generalized second law of thermodynamics as well as the thermodynamics of viscous dark energy have been recently studied by Setare and Shikhi (2010a, 2010b). Motivated the situation discussed above, in this paper, we study the impact of the interaction between viscous DE and DM  on the equation of state (EoS) parameter of dark energy in the scope of anisotropic Bianchi type I universe. We consider an energy flow from DE to DM and parameterize the interaction by a constant $\sigma$ and viscosity by $\zeta_{0}$, then a detailed investigation of the cosmological implications of this parametrization will be provided. As usual, a statefinder diagnostic will be performed in order to discriminate the different interaction parameters. Finally, by using $\chi^{2}$ method we investigate the constraints on the model parameters $\sigma$ and $H_{0}$ utilizing the recent observational data of Hubble rate $H(z)$.
%%%%%%%%%%%%%%%% %%%%%%%%%%%%%%%%%%%%%%%%%%%%%%%%%%%%%%%%%%%%%%%%%%%%%%%%%%%%%%%%%%%%%%%%%%%%%%%%%%%
%%%%%%%%%%%%%%%%%%%%%%%%%%%%%%%  SECTION 2  %%%%%%%%%%%%%%%%%%%%%%%%%%%%%%%%%%%%%%%%%%%%%%%%%%%%%%%
\section{The Metric and Field  Equations}
In an orthogonal form, the Bianchi type I line-element is given by
\begin{equation}
\label{eq1}
ds^{2} = -dt^{2} + A^{2}(t)dx^{2}+B^{2}(t)dy^{2}+C^{2}(t)dz^{2},
\end{equation}
where $A(t), B(t)$ and $C(t)$ are functions of time only. \\

The Einstein's field equations ( in gravitational units $8\pi G = c = 1 $) read as
\begin{equation}
\label{eq2} R^{i}_{j} - \frac{1}{2} R g^{i}_{j} = T^{(m) i}_{j} +
T^{(X) i}_{j},
\end{equation}
where $T^{(m) i}_{j}$ and $T^{(X) i}_{j}$ are the energy momentum tensors of dark matter and viscous dark energy,
respectively. These are given by
\[
  T^{m i}_{j} = \mbox{diag}[-\rho^{m}, p^{m}, p^{m}, p^{m}],
\]
\begin{equation}
\label{eq3} ~ ~ ~ ~ ~ ~ ~ ~  = \mbox{diag}[-1, \omega^{m}, \omega^{m}, \omega^{m}]\rho^{m},
\end{equation}
and
\[
 T^{X i}_{j} = \mbox{diag}[-\rho^{X}, p^{X}, p^{X}, p^{X}],
\]
\begin{equation}
\label{eq4} ~ ~ ~ ~ ~ ~ ~ ~ ~ ~ ~ ~ ~ ~ = \mbox{diag}[-1, \omega^{X}, \omega^{X},
\omega^{X}]\rho^{X},
\end{equation}
where $\rho^{m}$ and $p^{m}$ are the energy density and pressure of the perfect fluid
component while $\omega^{m} = p^{m}/\rho^{m}$ is its EoS parameter. Similarly,
$\rho^{X}$ and $p^{X}$ are, respectively the energy density and pressure of the viscous DE component while
$\omega^{X}= p^{X}/\rho^{X}$ is the corresponding EoS parameter. In this paper we assume the following expression for the pressure of the viscous fluid (Eckart 1940)
\begin{equation}
\label{eq5} p^{X}_{eff}=p^{X}+\Pi
\end{equation}
where $\Pi = -\xi(\rho^{X})u^{i}_{;i}$ is the viscous pressure and
$H = \frac{u^{i}_{;i}}{3}$ is the Hubble's parameter. Here $u^{i} = (1, 0, 0, 0)$ is the four velocity vector
satisfying $u^{i}u_{j} = -1$. It is important to note that as a consequence of the positive sign of the
entropy change in an irreversible process, $\xi$ has to be positive (Nojiri and Odintsov 2003).
In general, $\xi(\rho^{X})=\xi_{0}(\rho^{X})^{\tau}$, where
$\xi_{0}>0$ and $\tau$ are constant parameters.\\

In a co-moving coordinate system ($u^{i} = \delta^{i}_{0}$),
Einstein's field equations (\ref{eq2}) with (\ref{eq3}) and
(\ref{eq4}) for Bianchi type-I metric (\ref{eq1}) subsequently
lead to the following system of equations:
\begin{equation}
\label{eq6} \frac{\ddot{B}}{B}+\frac{\ddot{C}}{C}+\frac{\dot{B}\dot{C}}{BC}=-\omega^{m}\rho^{m}-\omega^{X}_{eff}\rho^{X}+\Pi,
\end{equation}
\begin{equation}
\label{eq7} \frac{\ddot{A}}{A}+\frac{\ddot{C}}{C}+\frac{\dot{A}\dot{C}}{AC}=-\omega^{m}\rho^{m}-\omega^{X}_{eff}\rho^{X}+\Pi,
\end{equation}
\begin{equation}
\label{eq8}\frac{\ddot{A}}{A}+\frac{\ddot{B}}{B}+\frac{\dot{A}\dot{B}}{AB}=-\omega^{m}\rho^{m}-\omega^{X}_{eff}\rho^{X}+\Pi,
\end{equation}
\begin{equation}
\label{eq9} \frac{\dot{A}\dot{B}}{AB}+\frac{\dot{A}\dot{C}}{AC}+\frac{\dot{B}\dot{C}}{BC}=\rho^{m}+\rho^{X}.
\end{equation}
Solving eqs. (\ref{eq6})-(\ref{eq9}), one can find (Saha 2005)
\begin{equation}
\label{eq10} A(t)=a_{1}a~ exp(b_{1}\int a^{-3}dt),
\end{equation}
\begin{equation}
\label{eq11} B(t)=a_{2}a~ exp(b_{2}\int a^{-3}dt),
\end{equation}
and
\begin{equation}
\label{eq12} C(t)=a_{3}a~ exp(b_{3}\int a^{-3}dt),
\end{equation}
where
\[
a_{1}a_{2}a_{3}=1,~~~~~~~b_{1}+b_{2}+b_{3}=0.
\]
Here $a=(ABC)^{\frac{1}{3}}$ is the average scale factor of Bianchi type I model. Using eqs. (\ref{eq10})-(\ref{eq12}) in eq. (\ref{eq9}) we obtain
\begin{equation}
\label{eq13} H^{2}=\frac{\rho^{m}+\rho^{X}}{3}+K a^{-6},
\end{equation}
which is the analogue of the Friedmann equation and $K=b_{1}b_{2}+b_{1}b_{3}+b_{2}b_{3}$, is a constant. Note that $K$ denotes the
deviation from isotropy e.g. $K=0$ represents flat FLRW universe.\\

Now we assume the following model-independent deceleration parameter (Xu et al. 2006)
\begin{equation}
\label{eq14}
q(z)=\frac{1}{2}-\frac{\alpha}{(1+z)^{\beta}},
\end{equation}
where $\alpha$ and $\beta$ are constants determined by the recent observational constraints.
From (\ref{eq14}) we observe that for $z\gg 1$, $q\to \frac{1}{2}$ which is corresponding
to matter dominated era whereas for $z=0$, the current value of deceleration parameter is obtained as
$q_{0}=\frac{1}{2}-\alpha$. In view of eq. (\ref{eq14}) and
\begin{equation}
\label{eq15}
q=\frac{1}{2}\frac{d \ln H^{2}}{d\ln (1+z)}-1,
\end{equation}
the Hubble parameter is obtained as
\begin{equation}
\label{eq16}
H(z)=H_{0}(1+z)^{\frac{3}{2}}\mbox{exp}\left[\alpha((1+z)^{-\beta}-1)/\beta\right].
\end{equation}
%%%%%%%%%%%%%%%%%%%%%%%%%%%%%%%%%%%%%%%%%%%%%%%%%%%%%%%%%%%%%%%%%%%%%%%%%%%%%%%%%%%%%%%%%%%%%%%%%%%%%%%
%%%%%%%%%%%%%%%%%%%%%%%%%%%%%%%%%%%%% SECTION 3 %%%%%%%%%%%%%%%%%%%%%%%%%%%%%%%%%%%%%%%%%%%%%%%%%%%%%%%%
\section{Interacting Viscous Dark Energy}
In this section we study the behavior of viscous dark energy when there is an energy flow from DE to dark matter (DM). In this case the law of energy-conservation equation ($T^{ij}_{;j} = 0$) which yields
\begin{equation}
\label{eq17} \dot{\rho}^{m} + 3\frac{\dot{a}}{a}(1 + \omega^{m})\rho^{m} +
\dot{\rho}^{X} +3\frac{\dot{a}}{a}(1 + \omega^{X})\rho^{X}= 0.
\end{equation}
can be written in the following form for barotropic and dark fluids
\begin{equation}
\label{eq18}\dot{\rho}^{m} + 3\frac{\dot{a}}{a}(1 + \omega^{m})\rho^{m}= Q,
\end{equation}
\begin{equation}
\label{eq19}\dot{\rho}^{X}+ 3\frac{\dot{a}}{a}(1 + \omega^{X})\rho^{X}= - Q,
\end{equation}
separately. Here the interaction between the dark components is shown by the quantity $Q$. To find a solution for the coincidence problem, we have to consider an energy transfer from the dark energy to the dark matter by assuming $Q > 0$. $Q > 0$, ensures that the second
law of thermodynamics is fulfilled (Pavon and Wang 2009). We note that from eqs. (\ref{eq18}) and (\ref{eq19}) it is easy to find that the unit of $Q$ may be of inverse of time. Hence, a first and natural candidate can be the Hubble factor
$H$ multiplied with the energy density. In this paper, following Amendola et
al. (2007) and Gou et al. (2007), we consider
%%%%%%%%%%%%%%%%%%%%%%%%%%%%%%%%%%%%%%%%%%%%%%%%%%%%%%%%%%%%%%%%%%%%%%%%%%%%%%%%%%%%%%%%%%%%
%%%%%%%%%%%%%%%%%%%%%%%%%%%%%%%%%%%%%%%%%%%%%%%%%%%%%% Figure 1 and 2 %%%%%%%%%%%%%%%%%%%%%%%%%%%
\begin{figure}[ht]
\begin{minipage}[b]{0.5\linewidth}
\centering
\includegraphics[width=\textwidth]{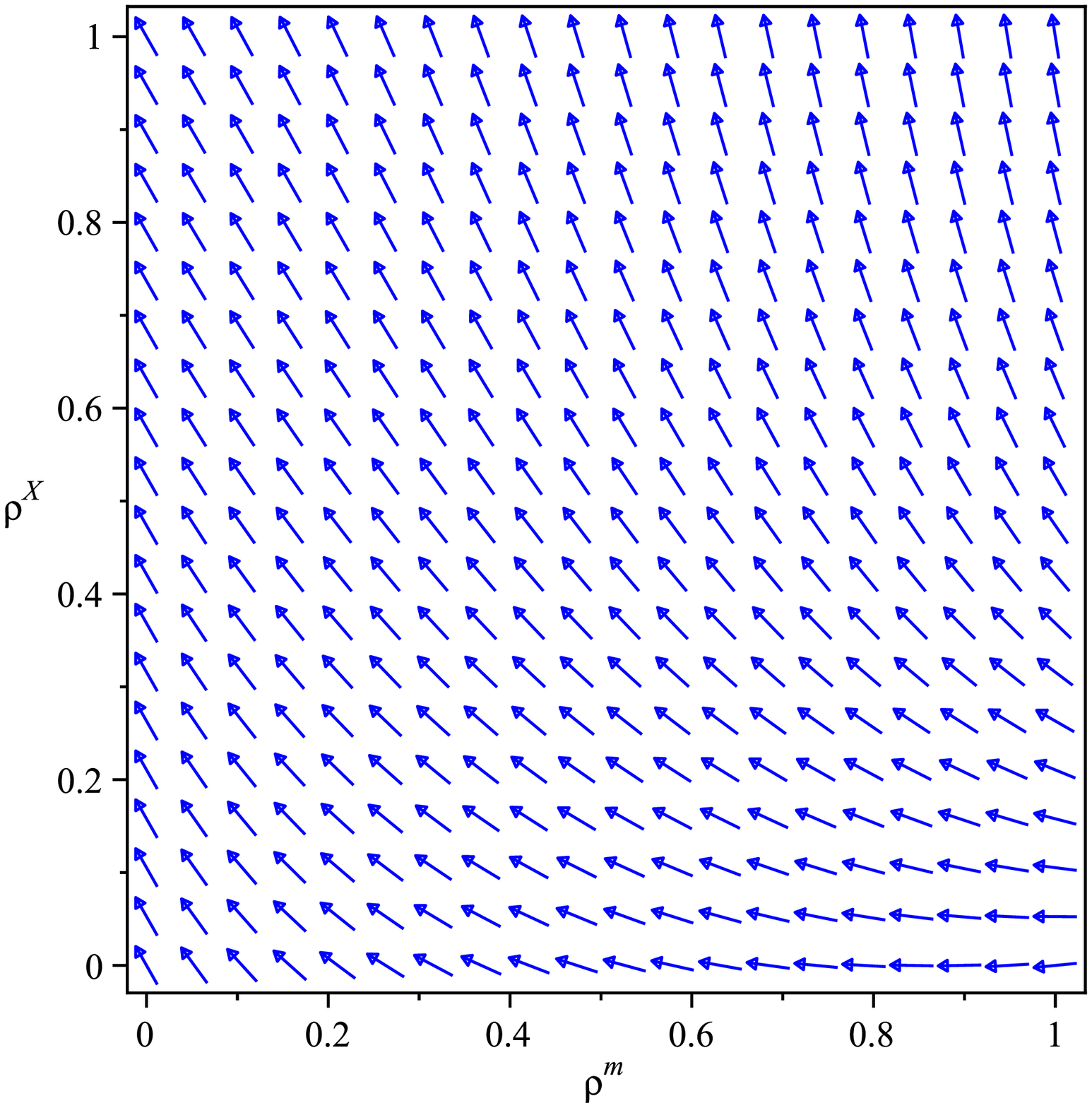} \\
\caption{Phase-plane diagram showing the evolution of the dark matter
density $\rho^{m}$ and the dark energy density $\rho^{X}$ for $\alpha =1.05$,
$\beta=1.43$, $H_{0}=72$, $\Omega^{m}_{0}=0.3$, $K=0.01$, $\omega^{m}=0$, and $\sigma=0$.
}
\end{minipage}
\hspace{0.5cm}
\begin{minipage}[b]{0.5\linewidth}
\centering
\includegraphics[width=\textwidth]{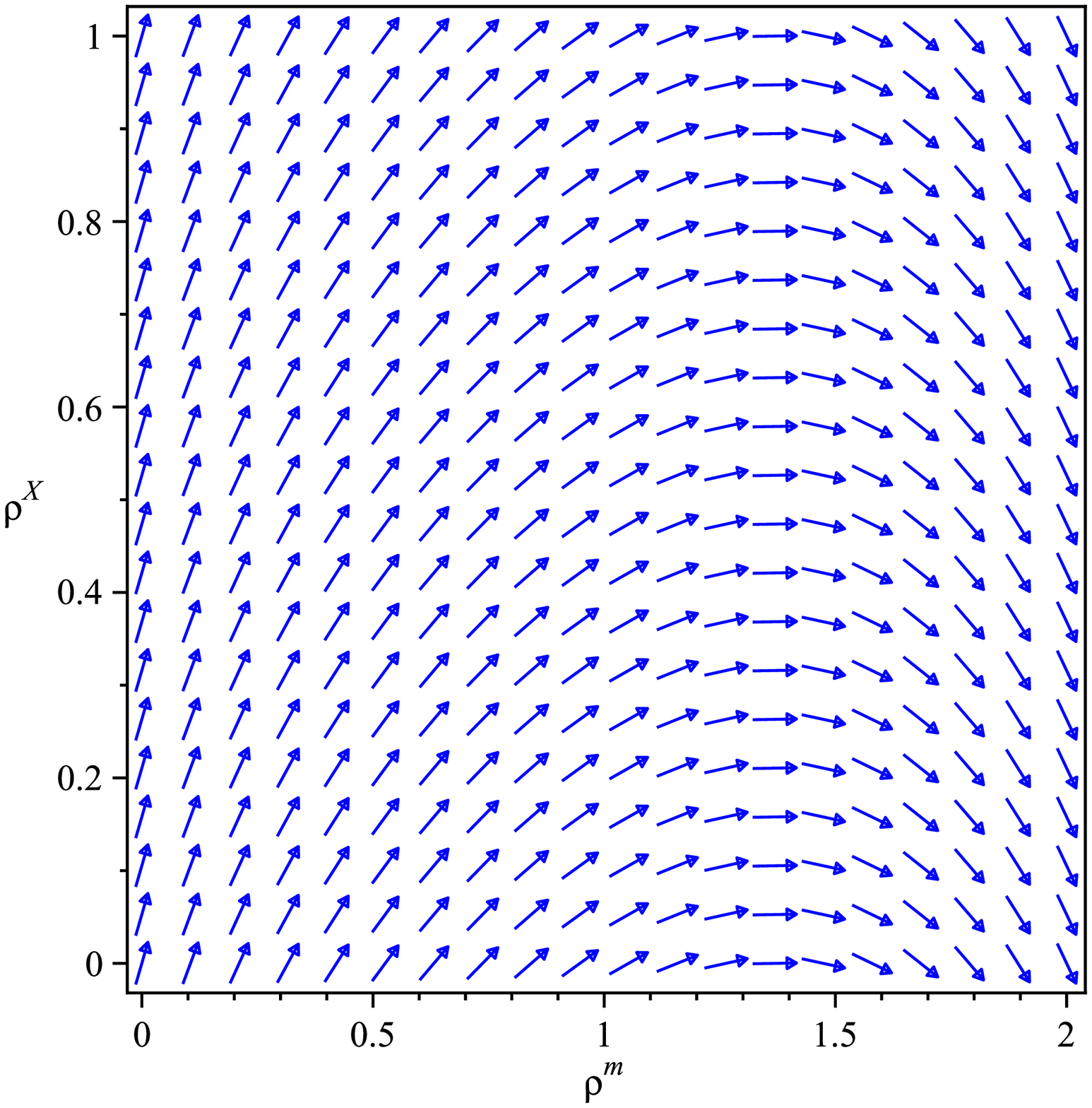}
\caption{Phase-plane diagram showing the evolution of the dark matter
density $\rho^{m}$ and the dark energy density $\rho^{X}$ for $\alpha =1.05$,
$\beta=1.43$, $H_{0}=72$, $\Omega^{m}_{0}=0.3$, $K=0.01$, $\omega^{m}=0$, and $\sigma=0.03$.}
\end{minipage}
\end{figure}
%%%%%%%%%%%%%%%%%%%%%%%%%%%%%%%%%% %%%%%%%%%%%%%%%%%%%%%%%%%%%%%%%%%%%%%%%%%%%%%%%%%%%%%%%%%%%
\begin{equation}
\label{eq20}Q =3 H \sigma \rho^{m},
\end{equation}
where $\sigma$ is a coupling coefficient which can be considered as a constant or a function of red shift. The combination of the SNLS, CMB, and BAO databases marginalized over a present dark energy density gives stringent constraints on the coupling, $-0.08 < \sigma < 0.03$ ($95\%$ C.L.) in the constant coupling model and $-0.4 < \sigma <0.1$ ($95\%$ C.L.) in the varying coupling model, where $\sigma$ is a present value (Guo et al. 2007).\\
We note that the eq. (\ref{eq20}) is not the only choose for $Q$ as one can select the following forms of $Q$: (i) $Q\propto H \rho^{X}$ and (ii) $Q\propto H (\rho^{m}+\rho^{X})$. The freedom of choosing $Q$ is because of our lack of knowledge of the nature of DE and DM.\\

Using eq. (\ref{eq20}) in eq. (\ref{eq18}) and after integrating, we obtain
\begin{equation}
\label{eq21}\rho^{m} = \rho_{0}^{m} a^{-3(1+\omega^{m}-\sigma)},
\end{equation}

where $\rho_{0}^{m}$ is an integrating constant. \\
By using eqs. (\ref{eq21}) and (\ref{eq16}) in eq. (\ref{eq13}), we obtain
\begin{equation}
\label{eq22} \rho^{X} =3 H^{2}_{0}\left((1+z)^{3}\mbox{exp}\left[\frac{2\alpha((1+z)^{-\beta}-1)}{\beta}\right]-\Omega_{0}^{m} (1+z)^{3(1+\omega^{m}-\sigma)}\right)-3K(1+z)^6,
\end{equation}
where we have used $a=(1+z)^{-1}$ and $\Omega_{0}^{m}=\frac{\rho^{m}_{0}}{3H^{2}}$.\\

Now, using eqs. (\ref{eq21}) and (\ref{eq22}) in eq. (\ref{eq6}), the effective EoS of dark energy $\omega^{X}_{eff}$ is obtained as
\[
\omega^{X}_{eff}=\frac{H^{2}(2q-1)+Ka^{-6}-\omega^{m}\rho^{m}}{\rho^{X}}-\xi_{0}(\rho^{X})^{\tau-1}
\]
\begin{equation}
\label{eq23} =-\frac{2 \alpha H^{2}_{0}(1+z)^{3-\beta}\mbox{exp}\left[\frac{2\alpha((1+z)^{-\beta}-1)}{\beta}\right]-K(1+z)^{6}+\omega^{m}\rho^{m}}{3 H^{2}_{0}\left((1+z)^{3}\mbox{exp}\left[\frac{2\alpha((1+z)^{-\beta}-1)}{\beta}\right]-\Omega_{0}^{m} (1+z)^{3(1+\omega^{m}-\sigma)}\right)-3K(1+z)^6}-\zeta_{0}H^{\tau}(\Omega^{X})^{\tau-1}.
\end{equation}
%%%%%%%%%%%%%%%%%%%%%%%%%%%%%%%%%%%%%%%%%%%%%%%%%%%%%%%%%%%%%%%%%%%%%%%%%%%%%%%%%%%%%%%%%%%%
%%%%%%%%%%%%%%%%%%%%%%%%%%%%%%%%%%%%%%%%%%%%%%%%%%%%%% Figure 3 and 4 %%%%%%%%%%%%%%%%%%%%%%%%%%%
\begin{figure}[ht]
\begin{minipage}[b]{0.5\linewidth}
\centering
\includegraphics[width=\textwidth]{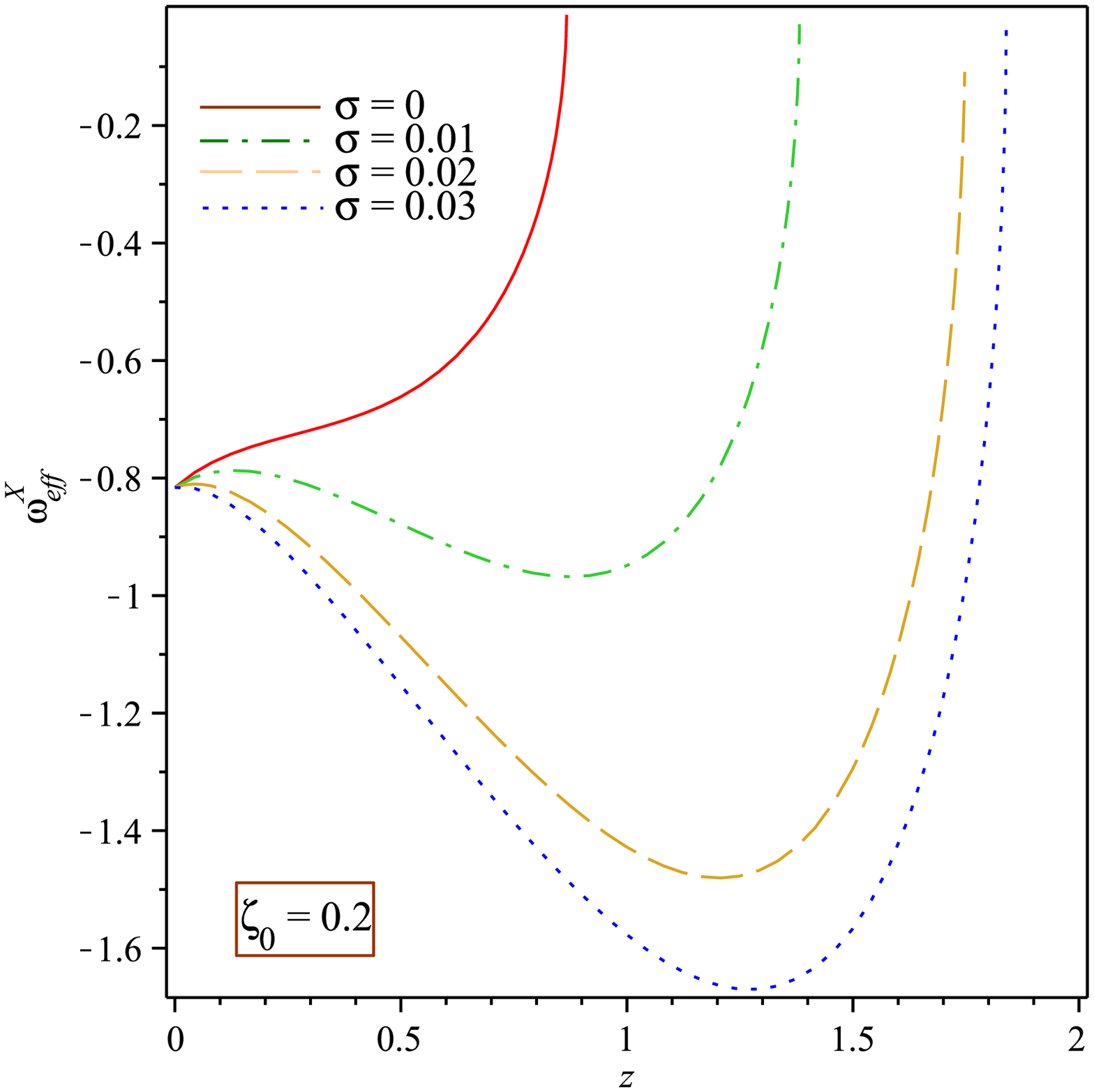} \\
\caption{The EoS parameter $\omega^{X}_{eff}$ versus $z$ for $\alpha =1.05$,
$\beta=1.43$, $H_{0}=72$, $\Omega^{m}_{0}=0.3$, $K=0.01$, and $\omega^{m}=0$. In this
case, we fix $\zeta_{0}=0.2$ and vary $\sigma$.}
\end{minipage}
\hspace{0.5cm}
\begin{minipage}[b]{0.5\linewidth}
\centering
\includegraphics[width=\textwidth]{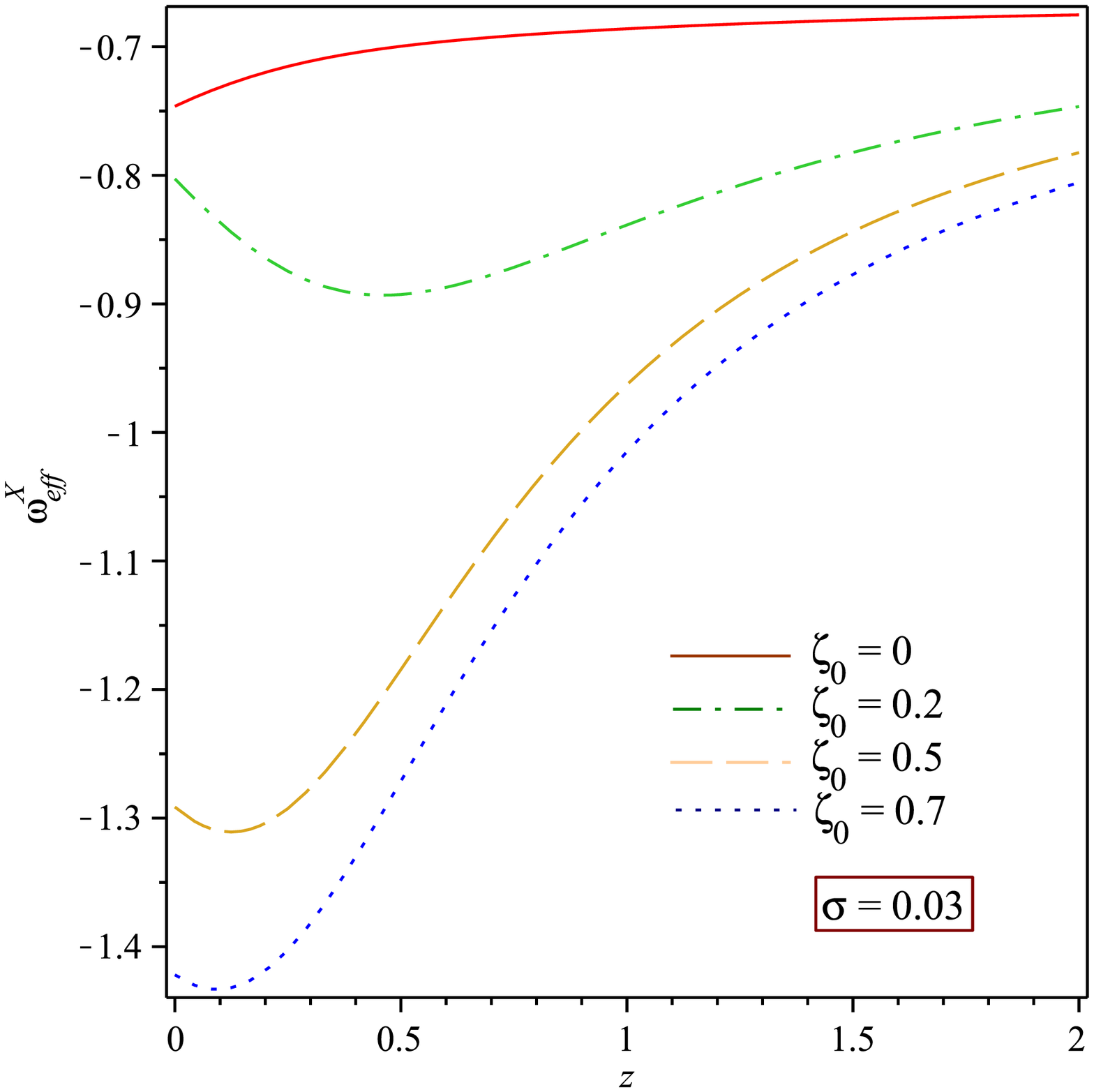}
\caption{The EoS parameter $\omega^{X}_{eff}$ versus $z$ for $\alpha =1.05$,
$\beta=1.43$, $H_{0}=72$, $\Omega^{m}_{0}=0.3$, $K=0.01$, and $\omega^{m}=0$. In this
case, we fix $\sigma=0.3$ and vary $\zeta_{0}$.}
\end{minipage}
\end{figure}
%%%%%%%%%%%%%%%%%%%%%%%%%%%%%%%%%%%%%%%%%%%%%%%%%%%%%%%%%%%%%%%%%%%%%%%%%%%%%%%%%%%%%%%%%%%%%
Note that here $\zeta_{0}=3^{\tau}\xi_{0}$, $q=-\frac{\frac{\ddot{a}}{a}}{H^{2}}$, and $\Omega^{X}=\frac{\rho^{X}}{3H^{2}}$.\\

Figures $1, 2$ show the Phase-plane diagram representing the evolution of the dark matter
density $\rho^{m}$ and the dark energy density $\rho^{X}$. In fig. 1 we have assumed that there is no interaction between dark components i.e $\sigma=0$. In this case the dark matter is continually being converted into the dark energy $\rho^{X}$, hence  as $\rho^{m}$
monotonically tends to $0$, $\rho^{X}$ tends monotonically to $\infty$. Therefore, the ratio $\frac{\rho^{X}}{\rho^{m}}$ always increases which
aggravates the coincidence problem. But as it is shown in fig. 2, interaction between matter and dark energy could lead to the energy transfer from dark energy to dark matter. The scenario of converting dark matter to dark energy and vis versa may occurs several times providing a cyclic decreasing-accelerating model of our universe. Such a model would provide a solution to the coincidence problem since the dark energy and dark matter will be comparable at infinitely several times.\\

The behavior of EoS ($\omega^{X}_{eff}$) parameter for dark energy in terms of red shift $z$ is depicted in Figures. $3, 4$. Since we are interested in the late time evolution of DE, we plot the range of red shift $z$ from $z=2$ to $z=0$ and the parameter $\omega^{m}$ is taken to be $0$. In Fig. $3$ we fix the parameter $\zeta_{0}=0$ and vary $\sigma$ as $0$, $0.01$, $0.02$, and $0.03$ respectively; in Fig. $4$ we fix $\sigma=0.03$ and vary
$\zeta_{0}$ as $0$, $0.2$, $0.5$, and $0.7$ respectively. The plots show that the evolution of $\omega^{X}_{eff}$ depends on the parameters $\sigma$ and $\zeta_{0}$ apparently. In another word, from fig. $3, 4$ we observe that the effective EoS parameter of an interacting viscous DE model varies from quintessence region to phantom region and ultimately approaches to the same value of $\omega^{X}_{eff}$ depending on the value of the bulk viscous coefficient $\zeta_{0}$. We also observe that either in non-interacting case with small $\zeta_{0}$ or interacting case with small $\sigma$, the EoS parameter only varies in quintessence region. It is worth to mention that while the current cosmological data from SNIa (Riess et al. 2004; Astier et al. 2006), CMB (Komatsu et al. 2009; Rodrigues 2008), and SDSS (Cruz et al. 2007) rule out that $\omega^{X}\ll -1$, they mildly favor a dynamically evolving DE crossing the PDE. Therefore, our results are in good agreement with well established theoretical and observational results.\\

The expressions for the matter-energy density $\Omega^{m}$ and dark-energy density $\Omega^{X}$ are given by
\begin{equation}
\label{eq24}
\Omega^{m} = \frac{\rho^{m}}{3H^{2}}=\Omega^{m}_{0}(1+z)^{3(1+\omega^{m}-\sigma)},
\end{equation}
%%%%%%%%%%%%%%%%%%%%%%%%%%%%%%%%%%%%%%%%%%%%%%%%%%%%%%%%%%%%%%%%%%%%%%%%%%%%%%%%%%%%%%%%%%%%
%%%%%%%%%%%%%%%%%%%%%%%%%%%%%%%%%%%%%%%%%%%%%%%%%%%%%% Figure 5 and 6 %%%%%%%%%%%%%%%%%%%%%%%%%%%
\begin{figure}[ht]
\begin{minipage}[b]{0.5\linewidth}
\centering
\includegraphics[width=\textwidth]{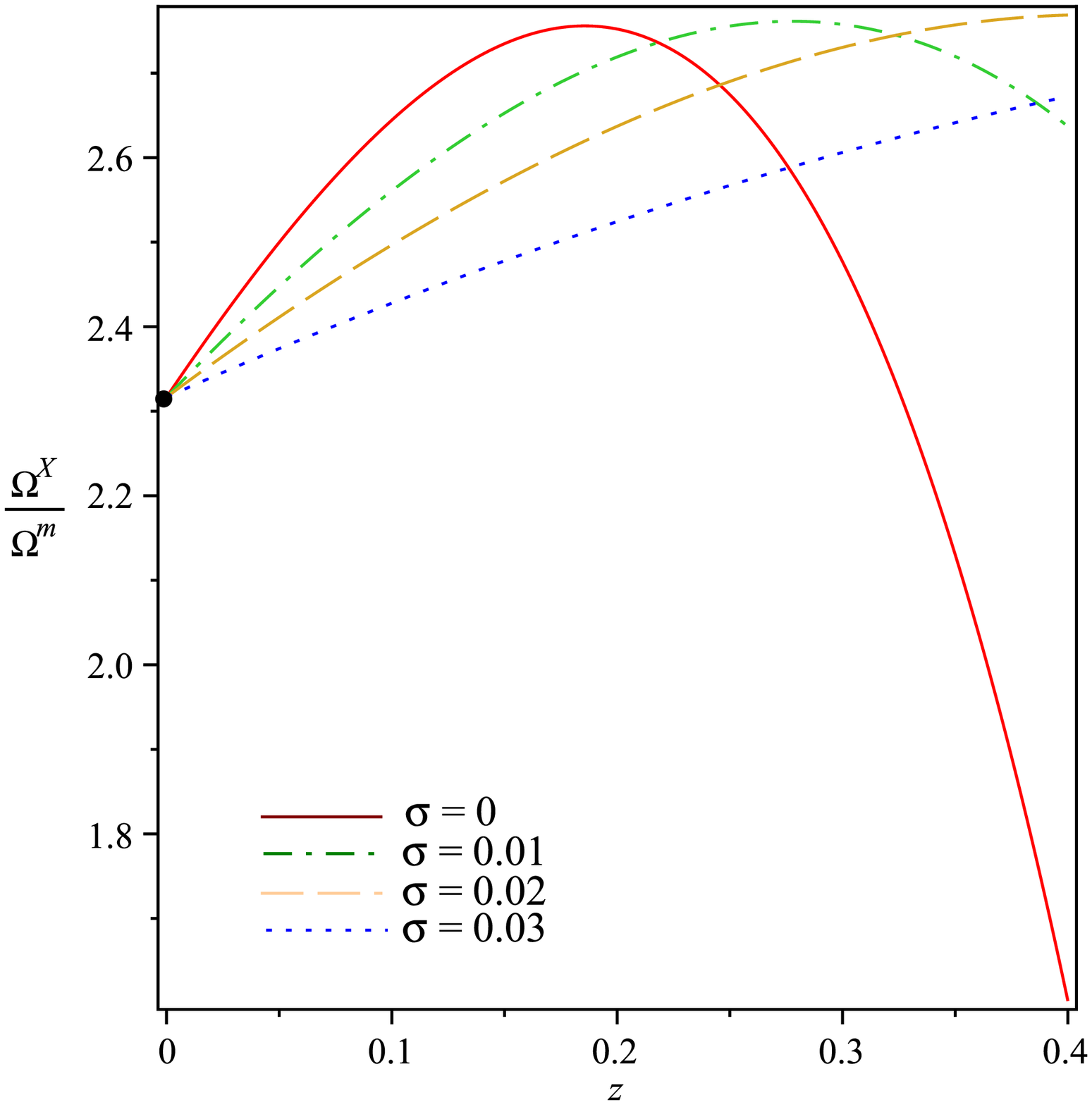} \\
\caption{The plot of the ratio of the DE and DM energy densities, $\frac{\Omega^{X}}{\Omega^{m}}$, versus red shift$z$ for $\alpha =1.05$,
$\beta=1.43$, $H_{0}=72$, $\Omega^{m}_{0}=0.3$, $K=0.01$, and $\omega^{m}=0$. The dot locate the
current value of this ratio.}
\end{minipage}
\hspace{0.5cm}
\begin{minipage}[b]{0.5\linewidth}
\centering
\includegraphics[width=\textwidth]{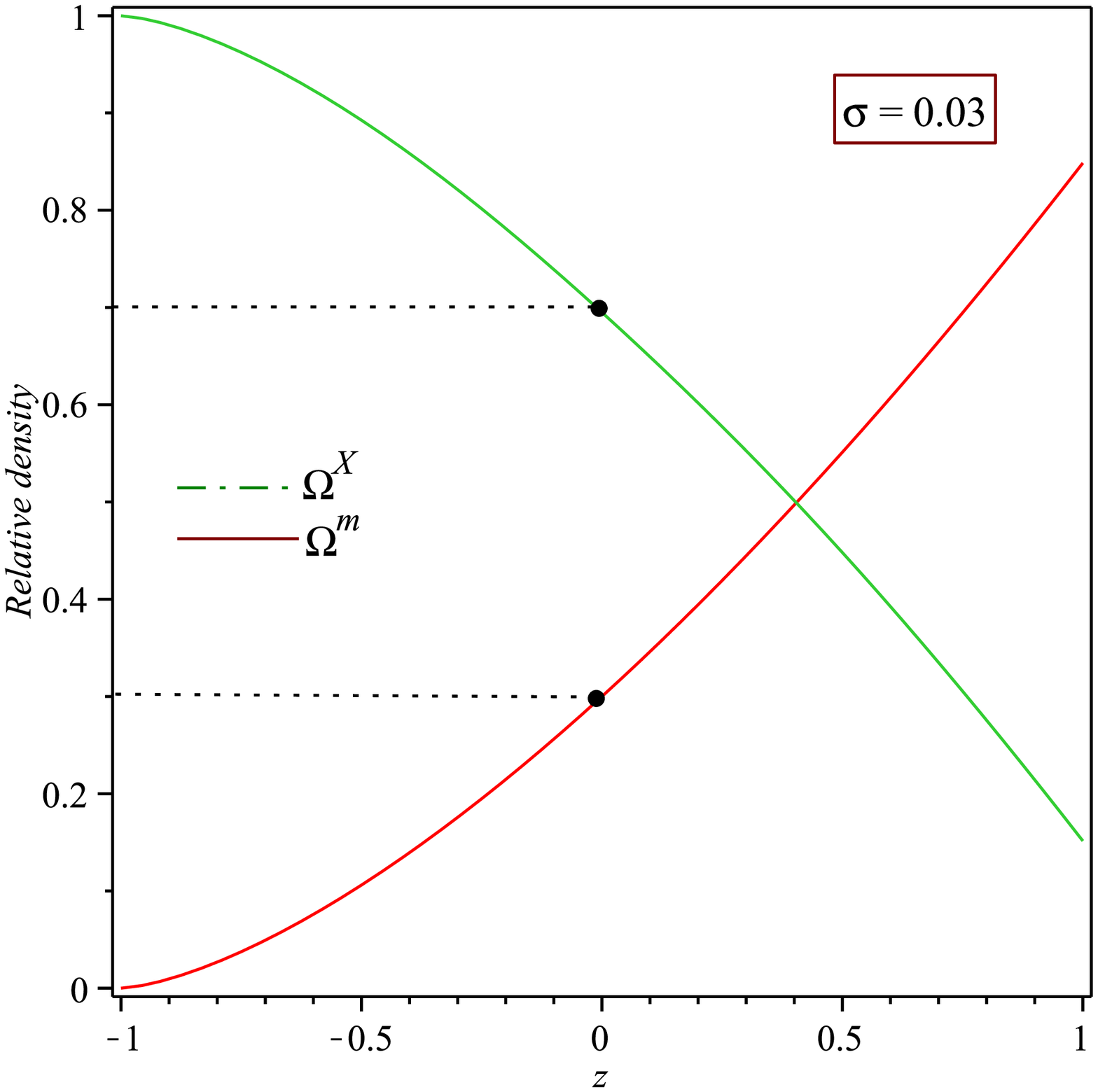}
\caption{The plot of $\Omega^{X}$ and $\Omega^{m}$ versus red shift $z$ $\alpha =1.05$,
$\beta=1.43$, $H_{0}=72$, $\Omega^{m}_{0}=0.3$, $K=0.01$, and $\omega^{m}=0$. The dots locate the
current values of $\Omega^{X}$ and $\Omega^{m}$. In this case, we fix $\sigma=0.03$.}
\end{minipage}
\end{figure}
%%%%%%%%%%%%%%%%%%%%%%%%%%%%%%%%%%%%%%%%%%%%%%%%%%%%%%%%%%%%%
and
\[
\Omega^{X} = \frac{\rho^{X}}{3H^{2}}=1-\Omega^{m}-\frac{Ka^{-6}}{3H^{2}}
\]
\begin{equation}
\label{eq25} =1-\Omega^{m}_{0}(1+z)^{3(1+\omega^{m}-\sigma)}-\frac{K(1+z)^{3}}{3H_{0}^{2}}\mbox{exp}\left[\frac{2\alpha(1-(1+z)^{-\beta})}{\beta}\right],
\end{equation}
respectively. \\
Adding Eqs. (\ref{eq24}) and (\ref{eq25}), we obtain
total energy ($\Omega$) as
\begin{equation}
\label{eq26}\Omega = \Omega^{m} + \Omega^{X}=1-\frac{K(1+z)^{3}}{3H_{0}^{2}}\mbox{exp}\left[\frac{2\alpha(1-(1+z)^{-\beta})}{\beta}\right].
\end{equation}
The variation of the ratio of DE and DM energy densities $\frac{\Omega^{X}}{\Omega^{m}}$ in our model is depicted in Fig. $5$. From this figure we observe that for high red shifts the DM is converting to DE whereas by decreasing red shift this conversion reverses due to the interaction between these two dark components. On other word, for small red shifts DE being converted to DM which leads to the solution of the coincidence problem. It is interesting to note that although the DE to DM energy transfer happens at smaller red shifts depending on the interaction strength, ultimately i. e at $z=0$, the $\frac{\Omega^{X}}{\Omega^{m}}$ ratio tends to the same valve independent to the interaction strength $\sigma$.
The permitted values of $\Omega^{m}$ and $\Omega^{X}$ in our model are also shown in Fig.$6$. Here, the dots locate the
current values of $\Omega^{X}$ and $\Omega^{m}$. From this figure we see that at late time the total energy density tends to one i.e $\Omega\to 1$. This result is compatible with the observational results. From both figures $5, 6$ we observe that the interaction parameter $\sigma$ brings impact on the evolution of the densities depending to its value.
%%%%%%%%%%%%%%%% %%%%%%%%%%%%%%%%%%%%%%%%%%%%%%%%%%%%%%%%%%%%%%%%%%%%%%%%%%%%%%%%%%%%%%%%%%%%%%%%%%%
%%%%%%%%%%%%%%%%%%%%%%%%%%%%%%%  SECTION 4  %%%%%%%%%%%%%%%%%%%%%%%%%%%%%%%%%%%%%%%%%%%%%%%%%%%%%%%
\section{Statefinder Diagnostic}
Since the investigation of dark energy, there have been many different DE models proposed to describe the current cosmic acceleration. To discriminating between these various contenders, Sahni et al (2003) introduced a new cosmological diagnostic pair $\{s, r\}$ called the statefinder in a model-independent manner. The parameters $s$ and $r$ are dimensionless and only depend on the scale factor $a$, therefore $\{s, r\}$ is a geometrical diagnostic. The above statefinder diagnostic pair has the following form
\begin{equation}
\label{eq27}
r\equiv\frac{\dot{\ddot{a}}}{aH^{3}},~~~~~s\equiv\frac{r-\Omega}{3(q-\frac{\Omega}{2})}.
\end{equation}
Here we extended the formalism of Sahni and coworkers to permit
curved universe models. In fact, trajectories in the $s-r$ plane corresponding to different cosmological models exhibit qualitatively different behaviors. For example, for quintessence and phantom models the trajectories lie in the region $s>0,~r<1$ whereas for
Chaplygin gas models trajectories lie in region $s<0,~r>1$. However, the quintessence, phantom and Chaplygin gas models tend
to approach the $\Lambda$CDM fixed point $\{s,r\}_{\Lambda CDM}=\{0,1\}$. \\
%%%%%%%%%%%%%%%%%%%%%%%%%%%%%%%%%%%%%%%%%%%%%%%%%%%%%%%%%%%%%%%%%%%%%%%%%%%%%%%%%%%%%%%%%%%%
%%%%%%%%%%%%%%%%%%%%%%%%%%%%%%%%%%%%%%%%%%%%%%%%%%%%%% Figure 3 and 4 Case 2 %%%%%%%%%%%%%%%%%%%%%%%%%%%
\begin{figure}[ht]
\begin{minipage}[b]{0.5\linewidth}
\centering
\includegraphics[width=\textwidth]{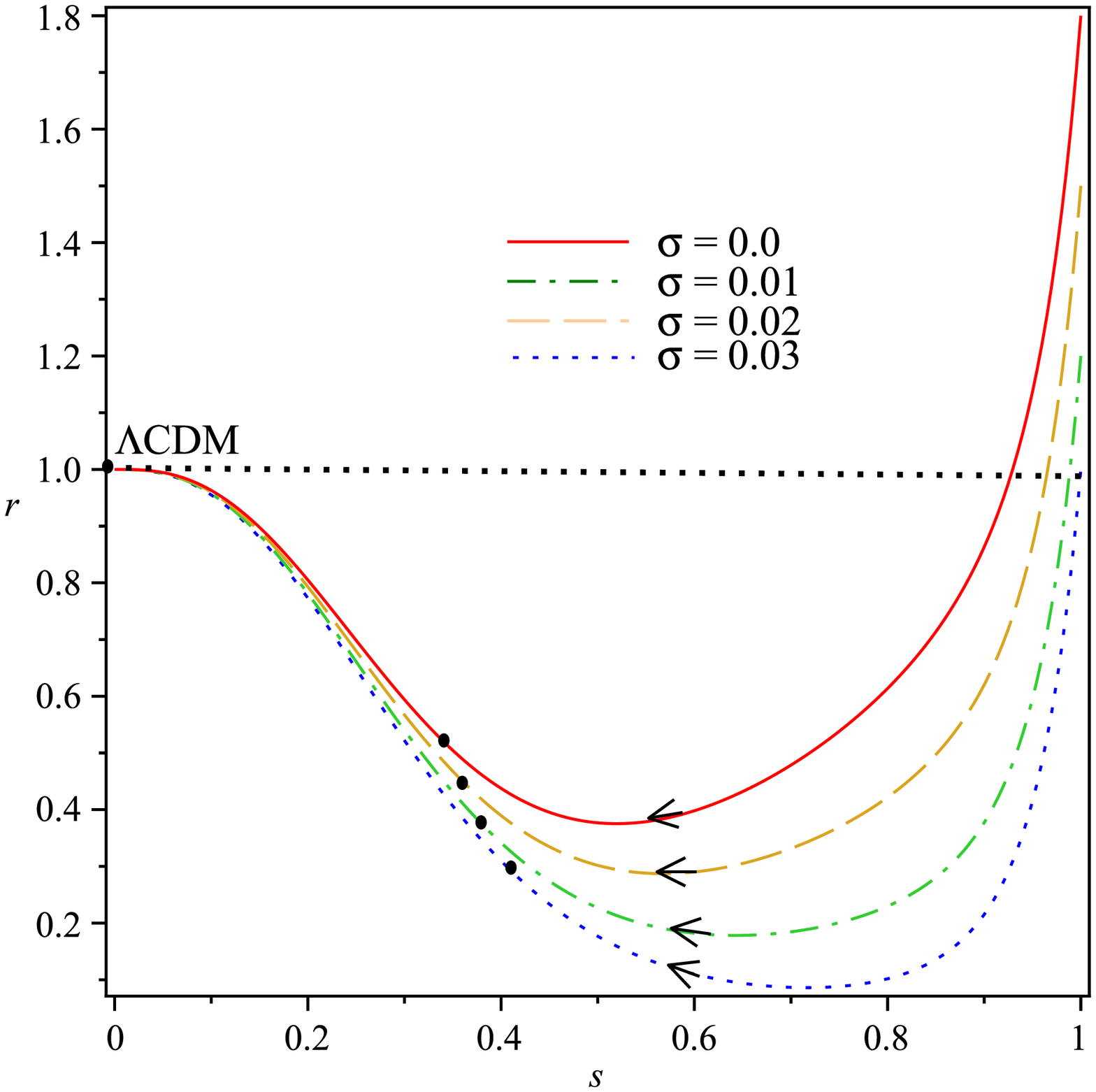} \\
\caption{$s-r$ evolution diagram. The dots locate the current
values of the statefinder pair $\{s,r\}$.}
\end{minipage}
\hspace{0.5cm}
\begin{minipage}[b]{0.5\linewidth}
\centering
\includegraphics[width=\textwidth]{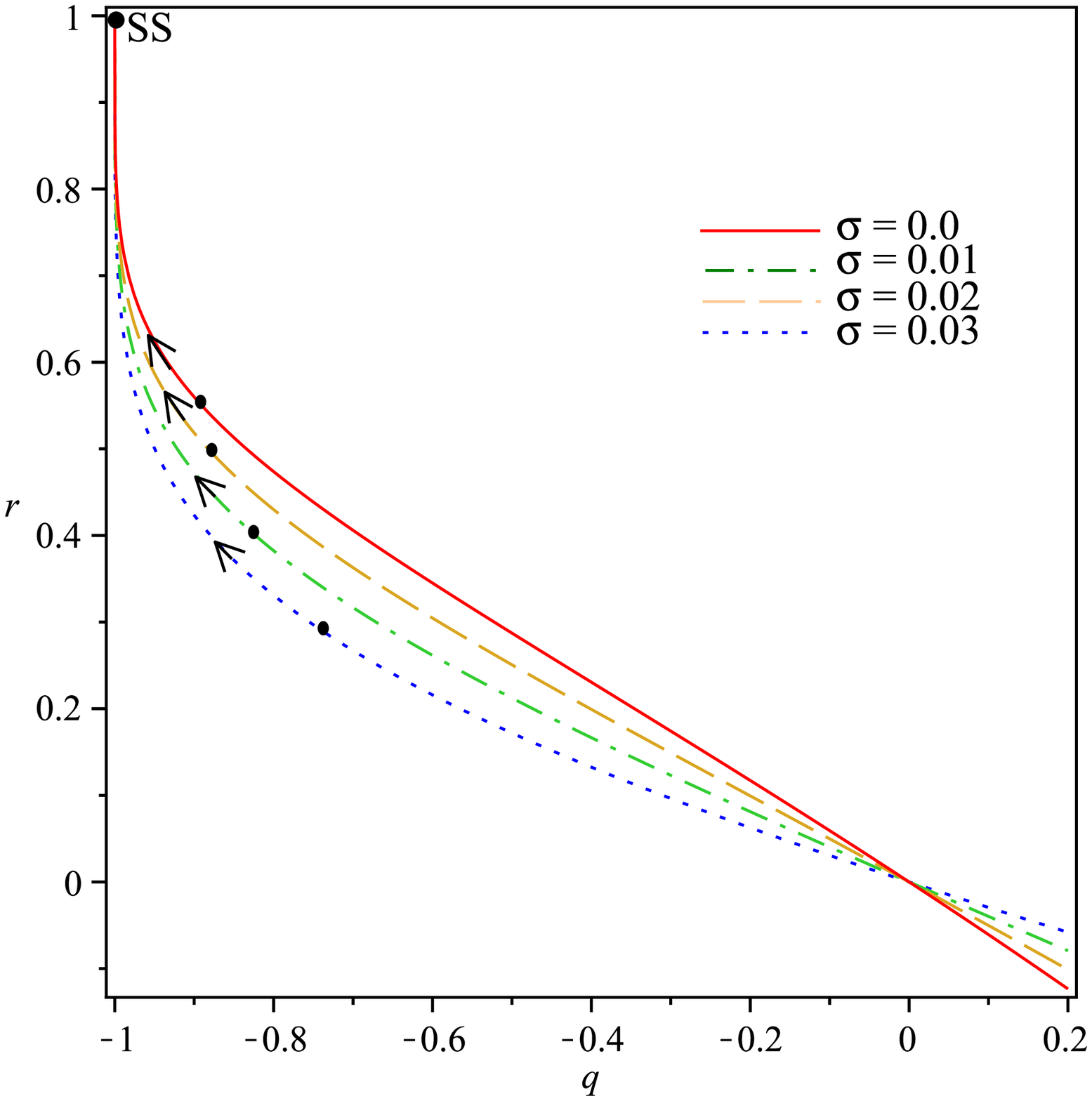}
\caption{$q-r$ evolution diagram. The dots locate the current
values of the pair $\{q,r\}$.}
\end{minipage}
\end{figure}
In general, the statefinder parameters are given by
\begin{equation}
\label{eq28}
r=\Omega^{m}+\frac{9\omega^{X}}{2}\Omega^{X}(1+\omega^{X})-\frac{3}{2}\Omega^{X}\frac{\dot{\omega}^{X}}{H},
\end{equation}
\begin{equation}
\label{eq29}
s=1+\omega^{X}-\frac{1}{3}\frac{\dot{\omega}^{X}}{\omega^{X}H}.
\end{equation}
Since from (\ref{eq23}), we have the analytical expression of $\omega^{X}_{eff}$ we can easily obtain
$\frac{\dot{\omega}^{X}_{eff}}{H}$. Thus, we can calculate the
statefinder parameters in this scenario.\\

The evolution diagrams for the statefinder pair $\{s,r\}$ and $\{q,r\}$ pair are shown in figures $7, 8$ respectively. The filled circles locate the current values of the pairs $\{s_{0},r_{0}\}$ and $\{q_{0},r_{0}\}$. In fig. $7$ we observe that $s$ decreases monotonically to zero, whereas $r$ first decreases to a minimum depending to the value of $\sigma$ and then rises to unity. In other word, our models tend to approach the $\Lambda CDM$ which is corresponding to the fixed point $(r=1, s=0)$ at $t\to\infty$.
Fig. $8$ shows that all models diverge at the same point in the past $(r=0, q=0)$ and converge to the same point in the future $(r=1, q=-1)$ which corresponds to the steady-state models (SS)- the de Sitter expansion ($\Lambda CDM\to SS$ as $t\to \infty$ and $\Omega^{m}=0$). From both figures $7, 8$, we observe that the interaction parameter $\sigma$ makes the model evolve along different trajectories on the $s-r$ and $q-r$
planes.
%%%%%%%%%%%%%%%% %%%%%%%%%%%%%%%%%%%%%%%%%%%%%%%%%%%%%%%%%%%%%%%%%%%%%%%%%%%%%%%%%%%%%%%%%%%%%%%%%%%
%%%%%%%%%%%%%%%%%%%%%%%%%%%%%%%  SECTION 5  %%%%%%%%%%%%%%%%%%%%%%%%%%%%%%%%%%%%%%%%%%%%%%%%%%%%%%%
\section{The experimental $H(z)$ test}
In this section, we use a direct fitting procedure involving the Hubble rate $H(z)$ to fix the cosmological bounds on $\sigma$ and $H_{0}$ in our model. To do this, we use the so called ``differential age" method proposed by Jimenez et al (2003). In this method we can directly study the observational $H(z)$ data without using the luminosity distance such as in the case of SNIa.\\

First of all, we note that the scaled Hubble parameter is given by
\begin{equation}
\label{eq30}
E(z)=\frac{H(z)}{H_{0}}=\left[\Omega^{m}(1+z)^{2}+\Omega^{X}\frac{\rho^{X}(z)}{\rho^{X}(0)}\right]^{\frac{1}{2}},
\end{equation}
where $\rho^{X}(z)$, $\Omega^{m}$, and $\Omega^{X}$ are given by eqs. (\ref{eq22}), (\ref{eq24}), and (\ref{eq25}).\\

Following Luongo (2011) we minimize the following reduced $\chi^{2}$ in order to constrain the model parameters.
\begin{equation}
\label{eq31}
\chi^{2}_{Hub}=\frac{1}{\nu}\sum_{i=1}^{14}\frac{[H^{th}(z_{i})-H^{obs}(z_{i})]^{2}}{\sigma^{2}_{Hub}(z_{i})},
\end{equation}
where $H^{obs}$ are the values of Table. 1 presented by Abraham et al. (2004) and $\nu = 13$. In eq. (\ref{eq31}), $H^{th}$ and $H^{obs}$ are referred to the theoretical and observational values for Hubble parameter respectively and the sum is taken over the cosmological dataset.\\
\begin{table}[ht]
\caption{The cosmological data at 1$\sigma$ error for $H(z)$ expressed in $s^{-1}MPc^{-1}Km$.}
\centering
\begin{tabular}{|ccccccccccccccc|}
\hline{\smallskip}
$z$ & 0.00 & 0.10 & 0.17 & 0.24 & 0.27 & 0.40 & 0.43 & 0.48 & 0.88 & 0.90 & 1.30 & 1.43 & 1.53 & 1.75 \\[0.5ex]
\hline
$H(z)$ & 72 & 69 & 83 & 79.69 & 77 & 95 & 86.45 & 97 & 90 & 117 & 168 & 177 & 140 & 220\\
$1\sigma~error$ & $\pm$8 & $\pm$12& $\pm$8 & $\pm$4.61 & $\pm$14 & $\pm$17& $\pm$5.96 & $\pm$60& $\pm$40& $\pm$23& $\pm$17 & $\pm$18 & $\pm$14 & $\pm$40 \\[1ex]
\hline
\end{tabular}
\label{table:nonlin}
\end{table}
%%%%%%%%%%%%%%%%%%% Figure 9 %%%%%%%%%%%%%%%%%%%%%%%%%%%%%%%%%%%%%%%%%%%%%%%%%%%%%%%%%%%%%%%%%%%%%%%%%%%%%
\begin{figure}[htbp]
\centering
\includegraphics[width=10cm,height=10cm,angle=0]{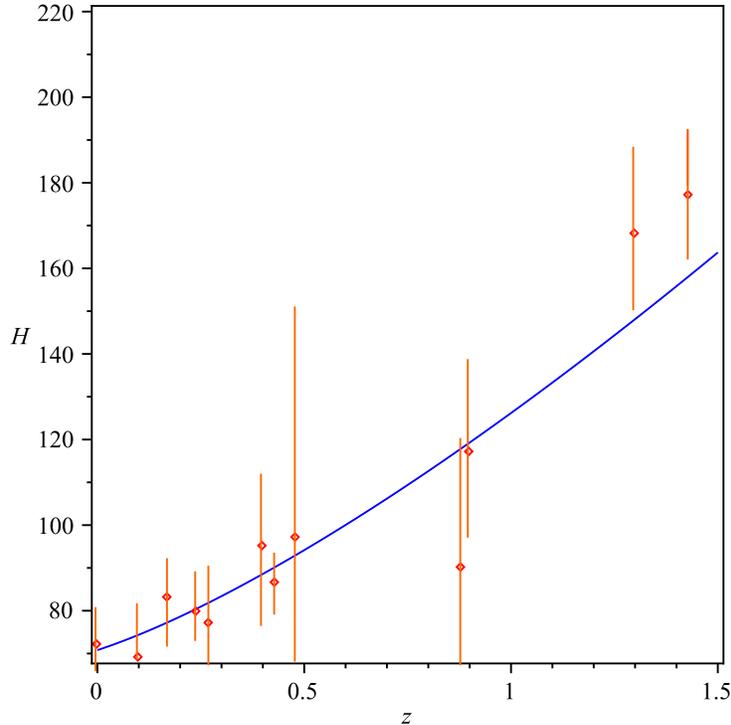}
\caption{The Hubble rate of our models versus the redshift $z$ for $\alpha=1.05$, $\beta=1.43$, and $H_{0}=70.8$. The points with bars
indicate the experimental data summarized in Table 1.}
\end{figure}
%%%%%%%%%%%%%%%%%%%%%%%%%%%%%%%%%%%%%%%%%%%%%%%%%%%%%%%%%%%%%%%%%%%%%%%%%%%%%%%%%%%%%%%%%%%%%%%%%

The constraints on model parameters $\sigma$ and $H_{0}$ we have to minimize $\chi^{2}_{Hub}$. To do so we fix all other parameters as follows: $\alpha=1.05$, $\beta=1.43$, $K=0.01$, $\Omega^{m}_{0}=0.3$, and $\omega^{m}=0$. Our results are given in Table. 2. As expected these results are in good agreement with the values obtained by Gou et al. (2007).\\
\begin{table}[ht]
\caption{The best fit parameters with 1$\sigma$ error.}
\centering
\begin{tabular}{|c|c|c|}
\hline{\smallskip}
Parameter & value & error\\[0.5ex]
\hline
$H(0)$ ($Km/s/M_{PC}$) & 70.8 & $\pm$1.30\\
\hline
$\sigma$ & 0.05 & $\pm$0.02\\
\hline
$\chi^{2}_{min}$ &3.29 &$\pm$1.65\\[1ex]
\hline
\end{tabular}
\label{table:nonlin}
\end{table}

The robustness of our fit can be viewed by looking at Fig. 9.
%%%%%%%%%%%%%%%%%%%%%%%%%%%%%%%%%% %%%%%%%%%%%%%%%%%%%%%%%%%%%%%%%%%%%%%%%%%%%%%%%%%%%%%%%%%%%
%%%%%%%%%%%%%%%%%%%%%%%%%%%%%%%% SECTION 8  %%%%%%%%%%%%%%%%%%%%%%%%%%%%%%%%%%%%%%%%%%%%%%%%%%%%%%%%%%%
\section{Concluding Remarks}
In this paper we have studied the interaction between viscous dark energy and dark matter in the scope of an anisotropic space-time. It is shown that in absence of interaction, dark matter continuously converts to dark energy whereas interaction leads to an energy transfer from DE to DM. Therefore, as expected, the interaction scenario could lead to the solution of the coincidence problem in cosmology. We found that either in non-interacting case with small $\zeta_{0}$ or interacting case with small $\sigma$, the EoS parameter only varies in quintessence region which is not in agreement with the recent observation. We have shown that to get an acceptable scenario of DE with an EoS parameter evolving from quintessence to phantom and then to the cosmological constant region ($\omega=-1$) one has to consider a viscous DE interacting with DM by choosing appropriate values for coupling constant $\sigma$ and the bulk viscous coefficient $\zeta_{0}$. In this case, our interacting viscous DE model does not encounter to the coincidence and big rip problems. Moreover, since in our model the viscosity dies out as the universe is expanding, the phantom phase is an temporary state and hence our model will not suffer from the ultraviolet quantum instabilities associated with the phantom models generated by the scalar fields. It is worth to mention that as argued by Carroll et al. (2003), any phantom model with $\omega<-1$ should decay to $\omega=-1$ at late time.
In addition, to differentiate between cosmological scenarios involving dark energy, a statefinder diagnostic has been preformed. This diagnostic indicate that the interaction between DE and DM can be probed by the statefinder parameters. Finally, we constraint the model parameters $\sigma$ and $H_{0}$ with the observational data for Hubble parameter by using chi-squared statistical method. It is worth to mention that the same method can be used for other Bianchi type space-times for which we can find the Friedmann-like equations. The study of interacting viscous DE in the scope of other anisotropic Bianchi type metrics will be published by the author soon.

%%%%%%%%%%%%%%%%%%%%%%%%%%%%%%%%%%%%%%%%%%%%%%%%%%%%%%%%%%%%%%%%%%%%%%%%%%%%%%%%%%%%%%%%%%%%%%%%%%%%%%%%%%
\section*{Acknowledgments}
This work has been supported by a research fund from the Mahshahr Branch of Islamic Azad University under the project entitled ``Interacting viscous dark energy and cold dark matter in an anisotropic universe". Author also would like to acknowledge the anonymous referee for fruitful comments.
%%%%%%%%%%%%%%%%%%%%%%%%%%%%%%%%%%%%%%%%%%%%%%%%%%%%%%%%%%%%%%%%%%%%%%%%%%%%%%%%%%%%%%%%%%%%%%%%%%%%%%%%%%%%

\end{document}